\documentclass{article}

\usepackage{microtype}
\usepackage{graphicx}
\usepackage{subfigure}
\usepackage{booktabs} 
\usepackage{graphicx}
\graphicspath{ {./images/} }

\usepackage{hyperref}

\usepackage[accepted]{icml2021}

\icmltitlerunning{UI Layout Generation with LLMs Guided by UI Grammar}

\begin{document}

\twocolumn[
\icmltitle{Exploring Mobile UI Layout Generation \\ using Large Language Models Guided by UI Grammar}



\icmlsetsymbol{equal}{*}

\begin{icmlauthorlist}
\icmlauthor{Yuwen Lu}{nd}
\icmlauthor{Ziang Tong}{nd}
\icmlauthor{Qinyi Zhao}{fdu}
\icmlauthor{Chengzhi Zhang}{cmu}
\icmlauthor{Toby Jia-Jun Li}{nd}
\end{icmlauthorlist}

\icmlaffiliation{nd}{University of Notre Dame, Notre Dame, USA}
\icmlaffiliation{fdu}{Fudan University, Shanghai, China}
\icmlaffiliation{cmu}{Carnegie Mellon University, Pittsburgh, USA}

\icmlcorrespondingauthor{Yuwen Lu}{ylu23@nd.edu}

\icmlkeywords{Human-Computer Interaction, User Interface, Large Language Models}

\vskip 0.3in
]



\printAffiliationsAndNotice{}  

\begin{abstract}
The recent advances in Large Language Models (LLMs) have stimulated interest among researchers and industry professionals, particularly in their application to tasks concerning mobile user interfaces (UIs). This position paper investigates the use of LLMs for UI layout generation. Central to our exploration is the introduction of \textit{UI grammar} –– a novel approach we proposed to represent the hierarchical structure inherent in UI screens. The aim of this approach is to guide the generative capacities of LLMs more effectively and improve the explainability and controllability of the process. Initial experiments conducted with GPT-4 showed the promising capability of LLMs to produce high-quality user interfaces via in-context learning. Furthermore, our preliminary comparative study suggested the potential of the grammar-based approach in improving the quality of generative results in specific aspects.
\end{abstract}

\section{Introduction}
\label{introduction}

\subsection{Mobile UI Layout Generation}
Layout generation for User interfaces (UIs), or Graphical User Interfaces (GUIs), has been explored by researchers across AI and Human-Computer Interaction (HCI). From a machine-learning perspective, the inherent multi-modal characteristics of UIs pose interesting research challenges for effective UI modeling, understanding, and generation~\cite{jiang2023future, jiang2022computational}; from an HCI perspective, UIs have been intensively studied as a medium for good user experience (UX). Various needfinding \cite{dow2005wizard, zimmerman2017speed, martelaro2017woz} and usability study \cite{nielsen1994usability, nielsen2005ten} methodologies have been developed both in academia and industry to improve the usability, functionality, and user-friendliness of UIs. Solving these challenges is seen as an early step to improving user experience at scale and reducing the workload for UI/UX designers \cite{lu2022bridging, knearem2023exploring}.

Following the release of the large-scale mobile UI dataset \textsc{RICO} \cite{deka2017rico}, several AI model architectures for mobile UI layout generation have been proposed. These architectures include but not limit to Generative Adversarial Network (GAN) \cite{li2019layoutgan, Kikuchi_Simo-Serra_Otani_Yamaguchi_2021}, Variational Autoencoder (VAE) \cite{Arroyo_Postels_Tombari_2021, Jing_Zhou_Tsang_Chen_Sun_Zhen_Du_2023}, Diffusion Model \cite{Cheng_Huang_Li_Li_2023, Hui_Zhang_Zhang_Xie_Wang_Lu_2023}, Graph Neural Network (GNN) \cite{Lee_Jiang_Essa_Le_Gong_Yang_Yang_2020}, and other Transformer-based neural networks \cite{Li_Amelot_Zhou_Bengio_Si_2020, Gupta_Lazarow_Achille_Davis_Mahadevan_Shrivastava_2021, Huang_Li_Zhou_Canny_Li_2021, kong2022blt, Sobolevsky_Bilodeau_Cheng_Guo_2023}.

\subsection{Large Language Models for UI Tasks}
Recent research work has explored some of LLMs' abilities on various UI-related tasks. \citet{wang2023enabling} utilized Large Language Models (LLMs) to conduct 4 UI modeling tasks through in-context learning and chain-of-thought prompting. \citet{liu2023chatting} conducted automated GUI testing by simulating human-like interactions with GUIs using LLMs. \citet{kargaran2023menucraft} explored user interface menu design with LLMs through natural language descriptions of designers' intentions and design goals. These efforts have demonstrated LLMs' capabilities to effectively work with UIs with careful interaction design and prompting techniques. Some experiments also exhibited competitive performance on UI task evaluation metrics, without the need for large-scale datasets or extensive training processes.




\subsection{Research problem and objectives}

In this work, we seek to explore LLMs' potential for generating mobile UI layouts. Specifically, we set out to determine how the in-context learning capabilities of LLMs can be harnessed in a one-shot learning scenario to generate high-quality UI layouts. A key challenge here involves the representation and integration of the hierarchical structure inherent in UI elements into the generation process.

\clearpage
In response to this problem, we propose \textit{UI grammar}—a novel approach that accurately represents the hierarchical relationships between UI elements. This approach serves to guide the generation process of LLMs, thereby making the generation more structured and contextually appropriate. From a human-centered perspective, we discuss how the inclusion of \textit{UI grammar} provides an intermediary layer of representation that could potentially improve the \textbf{\textit{explainability}} and \textbf{\textit{controllability}} of LLMs in the generation process. Users can better \textit{understand} and \textit{steer} LLMs' internal generation mechanisms by reviewing and editing the grammar used for coming up with the final result.

Our objectives here are twofold. First, we aim to evaluate the performance of LLMs in generating UI layouts. Second, we set out to evaluate the impact of our proposed \textit{UI grammar} on LLMs' generation process. We comparatively assessed the generation quality with/without the integration of \textit{UI grammar} in prompts against 3 common metrics for layout generation tasks: Maximum intersection over union (MaxIoU), Alignment, and Overlap. Our preliminary experiment results demonstrated LLMs' ability to generate high-quality mobile UI layouts through in-context learning and showcased the usefulness of \textit{UI grammar} in improving certain aspects of generation quality.

\section{LLM Prompting for UI Layout Generation}
\label{proposedmethod}

\citet{wang2023enabling} have discussed a few key aspects in constructing LLM prompts for mobile UI tasks, including \textit{screen representation}, \textit{UI element properties}, and \textit{class mappings}. While prompting LLMs remain an open research area, here, we continue this line of discussion by reviewing techniques from recent work on adapting UI for authoring LLM prompts. We then propose our own prompt strategy, specifically designed for UI layout generation, and provide our rationales.

\subsection{UI Representation for In-Context Learning}
LLMs have showcased an impressive capacity for in-context learning, which involves adapting to a limited number of user-provided examples while maintaining competitive performance across a variety of tasks \cite{brown2020language}. This ability, confirmed to be an emergent ability as language models' sizes scale up \cite{wei2022emergent}, offers a more streamlined alternative to the process of fine-tuning pre-trained models with large datasets for domain adaptation. \looseness=-1

UI data is inherently multi-modal and can often be represented in a variety of data formats. These include, but are not limited to, screenshots, Android view hierarchies, code implementations, and natural language descriptions (e.g., \textit{``a welcome page for a comics reading app''}). Within existing UI datasets, such as \textsc{RICO} \cite{deka2017rico}, each UI screen is typically offered in multiple data formats. This approach serves to capture visual, structural, and contextual information of UI screens.

This created challenges in providing UI exemplars to LLMs for in-context learning, especially given the limited context window and text-only input/output modality for existing LLMs. Here, we review recent work's approaches to adapting UI input for LLM prompting:

\begin{itemize}
    \item \citet{wang2023enabling} parsed UI into \textbf{HTML} files to feed into \textsc{PALM} for 4 mobile UI tasks, i.e. screen question-generation, screen summarization, screen question-answering, and mapping instruction to UI action. They used the \texttt{class}, \texttt{text}, \texttt{source\_id}, \texttt{content\_desc} attributes to include detailed information of screen widgets.
    
    \item \citet{liu2023chatting} investigated using GPT-3 for automatic GUI testing through natural language conversations.  They extracted static contexts using attributes of the app and screen widgets from the corresponding \textit{AndroidManifast.xml} file, including \texttt{AppName}, \texttt{ActivityName}, \texttt{WidgetText}, and \texttt{WidgetID}, and constructed \textbf{natural language sentences} describing the UI state with these attributes.
    
    \item While
 \citet{Feng_Zhu_Fu_Jampani_Akula_He_Basu_Wang_Wang_2023} did not directly work with UI data, they used GPT-3.5 for a 2D image layout generation, a task sharing many similarities with mobile UI layout generation. They parsed the position of image elements into \textbf{CSS} (short for Cascading Style Sheets) snippets with normalized position values as GPT input.
\end{itemize}

\subsection{Hierarchical Structures as UI Grammar}

UI elements within a screen have hierarchical relationships~\cite{li2021screen2vec,li2018appinite}, which can be reflected in the atomic design principle \cite{frost2016atomic}, the grouping feature of UI design tools like Figma, and the Android view hierarchies. Some previous work \cite{Huang_Li_Zhou_Canny_Li_2021} flattened the hierarchical structures of UI elements and reduced the layout generation task into predicting a flattened sequence of elements and the accompanying bounding boxes. However, our assumption is that preserving the hierarchical relationship between UI elements and using them to implicitly guide the generation process can improve the generation quality.

To include such hierarchical information into our prompt to guide LLMs in generation, here we take inspiration from previous work~\cite{kong2008adaptive, talton2012learning} and define \textbf{UI Grammar} as one possible way to represent the hierarchical relationship between UI elements. 

\textbf{UI Grammar} is defined as a set of production rules for describing the parent-children relationships between UI elements within a given screen hierarchical tree structure. Each production rule is of the form A $\rightarrow$ B, where A represents a parent UI element and B represents a sequence of one or more child elements. The definition resembles context-free grammar in syntax analysis \cite{earley1970efficient}, hence the name \textit{UI Grammar}.

\begin{figure}[htbp]
    \centering
    \includegraphics[width=\columnwidth]{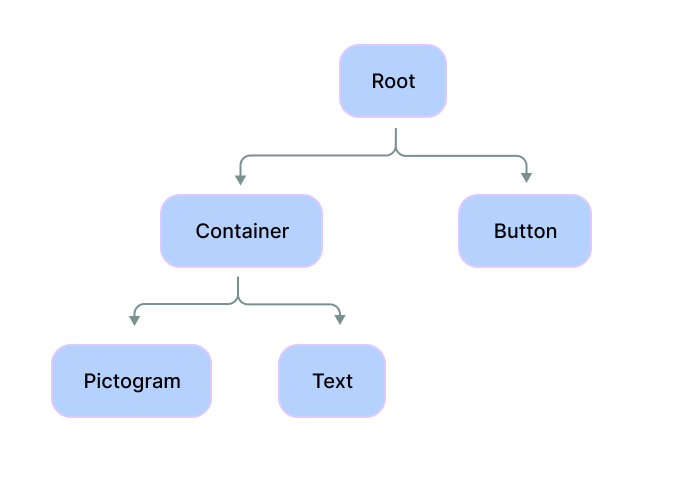}
    \caption{Example UI hierarchy structure. Here we can parse out UI grammar \texttt{Root $\rightarrow$ Container Button} and \texttt{Container $\rightarrow$ Pictogram Text}}
    \label{fig:example_tree}.
\end{figure}

For example, for a simple UI structure visualized in \ref{fig:example_tree}, we can parse out the following UI grammar based on the parent-children relationships: \texttt{Root $\rightarrow$ Container Button}, \texttt{Container $\rightarrow$ Pictogram Text}.

In this work, we conduct an initial comparative study between UI layout generation using LLMs \textit{with} and \textit{without} the guidance of UI grammar as part of the prompt.

\subsection{Problem Definition and Prompt Design}
We define our UI layout generation task as follows:

Given a natural language summary of a mobile UI screen $S$, we use LLM to generate a target hierarchical sequence of UI elements $T = \{o_{j} | j = 1, 2, \ldots, n_{u}\}$ where $o_{j}$ denotes a tuple of $(\texttt{label}, \texttt{bounding\_box})$ for UI element $j$. These two fields in the tuple represent the \textit{type} and \textit{position} of the UI element on the screen.

With this problem definition, we design our prompt with the following objectives:

\begin{enumerate}
    \item Using a UI format easy for LLMs to understand and generate layouts
    \item Encapsulating hierarchical relationship between UI elements through UI grammar
    \item Removing redundant non-visual information that is non-essential for layout generation
\end{enumerate}

Based on these objectives, we chose to use JSON as the data format to represent UIs in UI layout generation. We selected JSON due to its advantages in the following aspects:

\begin{itemize}
    \item \textbf{\textit{Compatibility:}} JSON is ideal and commonly used for data with hierarchically structured relationships. Also, given that many LLMs use programming code in training data and prompts falling within the training data distribution tend to perform better \cite{wang2023enabling}, JSON is a compatible data format on both ends for our task. \looseness=-1
    \item \textbf{\textit{Flexibility:}} JSON supports multiple types of attributes for each element, suitable for representing the string.\texttt{label} and the list of integer coordinates for \texttt{bounding\_box} \looseness=-1
    \item \textbf{\textit{Processing Simplicity:}} UI datasets such as \textsc{RICO} already use JSON to represent UI view hierarchy, reducing processing efforts.
\end{itemize}

In order to compare the efficacy of UI layout generation \textit{with} and \textit{without} the guidance from UI grammar, we have created two analogous pipelines for the generation process (Fig \ref{fig:pipeline1}, Fig \ref{fig:pipeline2}). The main differentiation between these pipelines lies in the inclusion of UI grammar in our prompts for LLMs. We will first discuss the pipeline that operates without the UI grammar, then introduce how we integrated UI grammar into the prompt to steer LLMs in generating UI layouts.

Rather than directly work with the \textsc{RICO} dataset containing approximately 66k unique UI screens, we used an improved dataset \textsc{Clay} \cite{li2022learning} that is based on \textsc{RICO}. \textsc{Clay} removed noise from \textsc{RICO} UI data by detecting UI element types and visual representation mismatches and assigning semantically meaningful types to each node. It contains 59k UI human-annotated screen layouts and contains less-noisy visual UI layout data. We also utilized the \textsc{Screen2Words} dataset \cite{wang2021screen2words} which contains natural language summaries of UI screens in \textsc{RICO} to construct our prompt.

\subsubsection{Prompt Without UI Grammar}

\begin{figure*}[t]
    \centering
    \includegraphics[width=\textwidth]{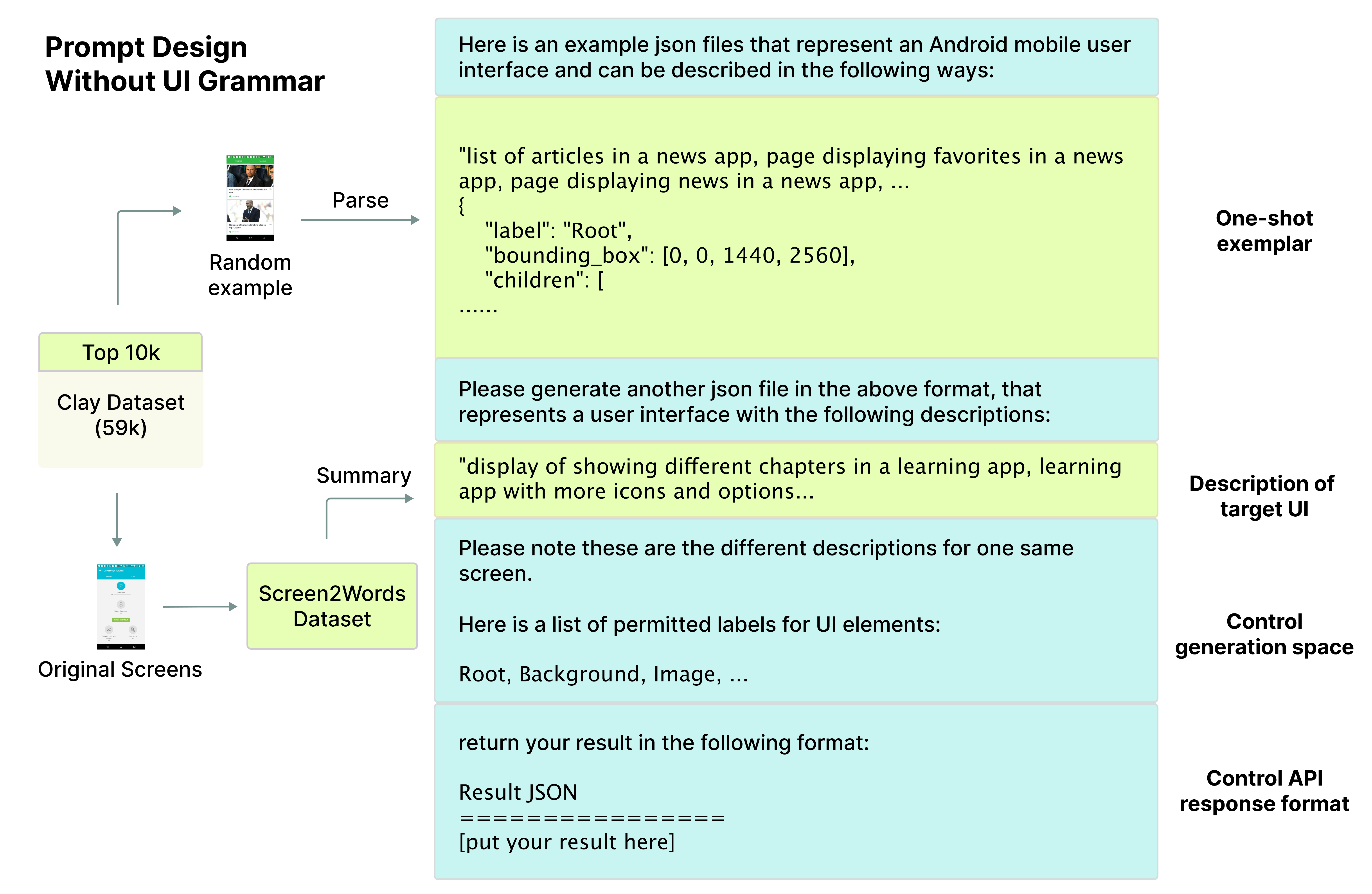}
    \caption{Prompt 1 design for generation without UI grammar}
    \label{fig:pipeline1}
\end{figure*}

For generating layouts without involving UI grammar (Fig. \ref{fig:pipeline1}), we first randomly select a screen from \textsc{Clay} to use as an example for our in-context learning (i.e. 1-shot prompting) and exclude it from generation to prevent data leakage. For each UI screen in \textsc{Clay}, we retrieve the corresponding natural language description from \textsc{Screen2Words} as the description for our generation target. To control the generation result and only receive layouts with meaningful UI elements, we used the semantically meaningful list of 25 UI element labels defined in \textsc{Clay} and included that in our prompt as a constraint. We also controlled LLM's API response format for easier parsing of the generation result, as shown in Fig \ref{fig:pipeline1}.  

\subsubsection{Prompt With UI Grammar}
\label{prompt_with_grammar}
\begin{figure*}[t]
    \centering
    \includegraphics[width=\textwidth]{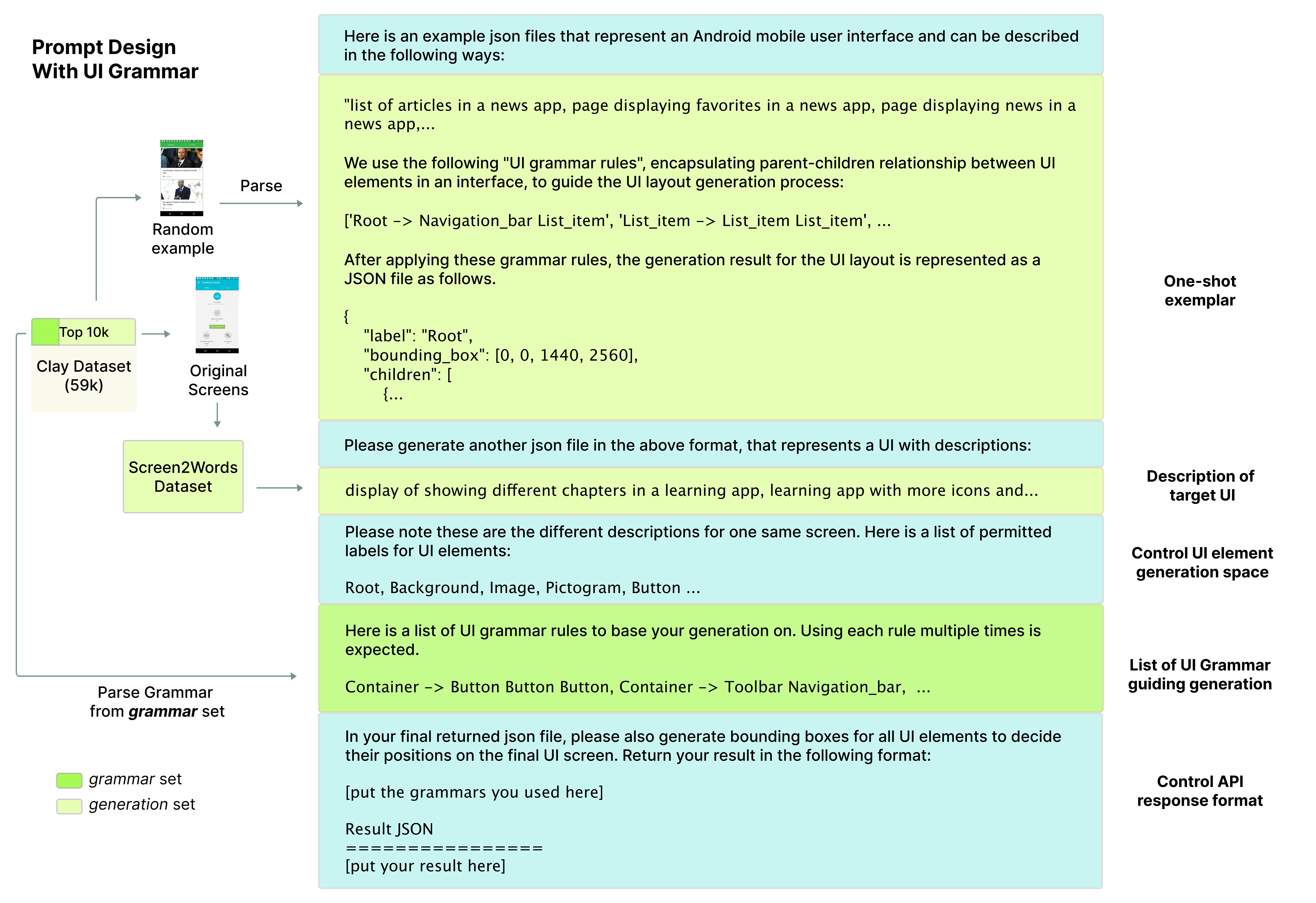}
    \caption{Prompt 2 design for generation with UI grammar}
    \label{fig:pipeline2}
\end{figure*}

For our second pipeline and prompt design (Fig. \ref{fig:pipeline2}), we introduce UI grammar as an intermediary step in UI layout generation with an architecture similar to neuro-symbolic models \cite{sarker2021neuro}. Instead of asking LLMs to directly generate the final screen layout, in the 1-shot example, we describe UI layout generation as a 2-step process: first, we introduce the list of UI grammar in the screen, then explain how we can generate the example UI layout using the provided UI grammar.

An important step in constructing the prompt with UI grammar is selecting which screens from the \textsc{Clay} dataset to parse grammar from. When generating a layout using descriptions of screen $S$ from the original \textsc{Clay} dataset, if we also input grammars parsed from $S$ into the prompt, data leakage occurs as screen $S$ can be reconstructed from its own grammars in a straightforward manner. To avoid this, we conduct a 20/80 random split of the \textsc{Clay} dataset and use grammars parsed from the 20\% grammar set to guide the generation of the 80\% generation set.

In addition, from our observation, many screens from the same app packages in \textsc{Clay} share similar layout structures. Consequently, we splitted the dataset by apps in order to avoid the data leak caused by having screens from the same app package in both sets.

\section{Initial Experiments}
\label{experiments}

In May 2023, we used OpenAI's GPT-4 API to conduct a preliminary experiment comparing the 2 proposed pipelines for UI layout generation. We used the \texttt{gpt-4-0314} version of GPT-4 with a \texttt{max\_token} of $2,000$ and \texttt{temperature} of $0.7$.

\textbf{Dataset} For both prompt designs, we pre-process the UI view hierarchy files from \textsc{Clay} by removing all attributes of UI elements but \texttt{label} and \texttt{bounds}, as all others are not necessary for our layout generation task. To further ensure the generation quality, we work with a subset of the top $10k$ UI screens from \textsc{Clay} with an app review higher than $4.3$ and download of more than $10k$ in Google Play Store for our generation. These two thresholds serve as quality filters and are manually defined to balance the need for a sufficiently large sample size against the desire for high-quality app representation.

\begin{figure*}[t]
    \centering
    \includegraphics[width=\textwidth]{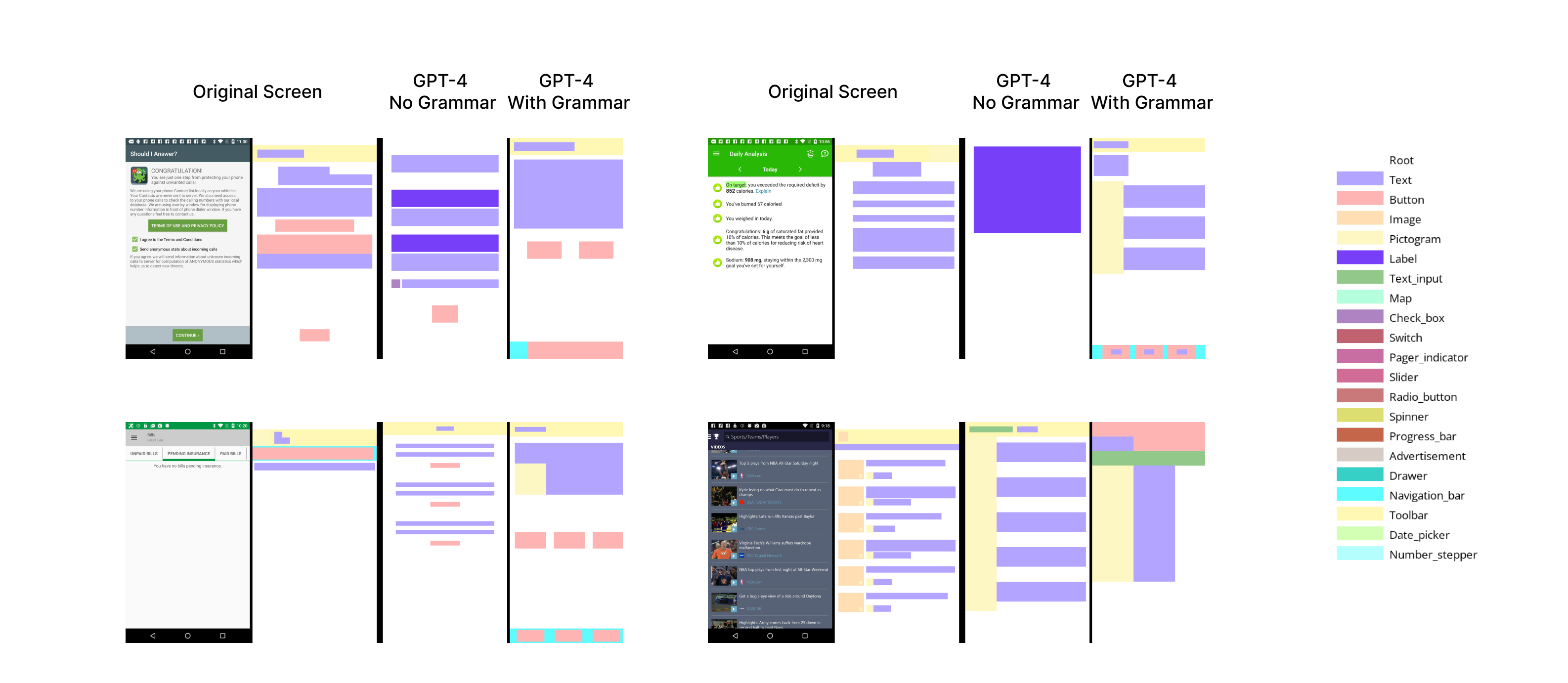}
    \caption{Visualizing the generation results. In each 4-screens group, the left 2 images are the original image and its parsed bounding boxes, while the right 2 are GPT-4 generated results. The original image's description from \textsc{Screen2Words} was used in the prompt for generating the right 2 layouts.}
    \label{comp_results}
\end{figure*}

Given OpenAI's API response rate and call limits, it is hard to quickly generate a large number of results. In this work-in-progress, we have conducted an initial experiment on a batch of $192$ UI screens from the top apps in \textsc{Clay} and report the preliminary evaluation results as follows. Visualization of example generation results is shown in \ref{comp_results}.

\section{Preliminary Evaluations}
\label{evaluations}

Here we report preliminary evaluations of our UI layout generation results against 3 common metrics commonly used in this domain: Maximum Intersection Over Union (MaxIoU), Alignment, and Overlap. \footnote{Refer to \citet{Jing_Zhou_Tsang_Chen_Sun_Zhen_Du_2023} for definitions of these metrics.} The MaxIoU value is calculated between the generation screen $S'$ and the original screen $S$ from \textsc{Clay} the provided screen summaries as part of the prompt. Alignment and Overlap are both calculated over the generated result $S$ only.

Please note that in order to more accurately evaluate the visual quality of the generated UI layouts, we removed 5 types of UI elements that are commonly invisible on screens \footnote{Namely: \textsc{Root}, \textsc{Background}, \textsc{List\_item}, \textsc{Card\_view}, and \textsc{Container}} from the results before evaluation. 

\textbf{Results} In Table \ref{tab:eval_table}, we can see that in our initial experiment, GPT-4 performed well on \textit{overlap} without grammar, and on \textit{alignment} with and without grammar, having close or even better metric performance than real data. The \textit{overlap} result for both prompt designs achieved $0.00$, meaning every element aligns with at least 1 other element on the screen. This is consistent with the visual appearance of the generation results. While we did not specifically mention the need to align UI elements or avoid element overlap in our prompt, GPT-4 was able to generate high-quality results against these metrics. 
In addition, introducing UI grammar to guide GPT-4's layout generation process slightly increased the MaxIoU performance. On this metric, GPT-4 with grammar is comparable with some general layout generation models trained on large datasets as reported in \cite{Jing_Zhou_Tsang_Chen_Sun_Zhen_Du_2023}, demonstrating LLMs' in-context learning ability in mobile UI layout generation.

While we did not explicitly restrict GPT-4 on using only the provided grammar set \footnote{Specifically, we used the wording ``Here is a list of UI grammar rules to base your generation on. Using each rule multiple times is expected''.}, $83.8\%$ of the rules GPT-4 reported to be using for generation came from the provided grammar. This showed that GPT-4 was not entirely restricted by the grammar we provided, demonstrating the flexibility of the model and our approach. \looseness=-1

\begin{table*}[ht]
\centering
\begin{tabular}{c|c|c|c}  
\hline  
 & MaxIoU $\uparrow$ & Overlap $\downarrow$ & Alignment $\downarrow$ \\  
\hline
GPT-4 no grammar & $0.29$ & $8.14$ & $0.00$ \\
\hline
GPT-4 with grammar & $0.34$ & $12.47$ & $0.00$ \\
\hline
Real Data & ---- & $8.58$ & $0.00$ \\
\hline
\end{tabular}
\caption{Quantitative comparisons of our 2 approaches and real data from \textsc{Clay}}
\label{tab:eval_table}  
\end{table*}

\section{Discussion and Future Work}
\label{discussion}

Our experimentation with LLMs for UI layout generation has demonstrated LLMs' promising ability on this task. However, we believe LLMs like GPT-4 also have the potential capability to generate content along with layouts to create mid-fi to high-fi prototypes. The potential to combine LLMs with existing UI templates or design systems such as Google Material Design will enable more automated, customized, and efficient UI prototyping techniques.

We argue that besides improving LLMs' generation quality on metrics like MaxIoU, by introducing UI grammar as an intermediary representation in the generation process, our approach could increase the explainability and users' controllability of black-box pre-trained LLMs:

\begin{itemize}
    \item \textbf{\textit{Explainability:}} By reviewing the UI grammar employed in LLMs' generation processes, users could gain a better understanding of LLMs' internal generation mechanisms. Our approach differs from a post-hoc explanation request for LLMs, in that our approach can be more easily verified through an easy comparison between the grammars employed and the final UI structure. On the other hand, post-hoc explanation requests (e.g. a follow-up question such as \textit{``explain why you generated this result''}), while similar to how humans provide justifications, do not necessarily reflect the actual generation mechanism.
    \item \textbf{\textit{Controllability:}} With UI grammar as an intermediary representation in the generation process, users can obtain higher control of the generation results if enabled to modify or replace the grammar provided to the LLMs in prompts. Future applications can build upon such model architecture and provide users with more ways to interact with UI grammar in the prompts (e.g. directly selecting which apps to extract grammar from) to improve the controllability of LLMs in similar generation tasks.
\end{itemize}

In Section~\ref{prompt_with_grammar} we have discussed the potential of data leak when using the \textit{natural language description} and \textit{UI grammar} derived from the same screen. But on the other hand, since UI grammar represents different ways of organizing and designing UI elements on a screen, we could potentially use UI grammar as a proxy to control characteristics of generation results. One possible use case is generating certain styles of UI, by extracting grammar specifically from screens in compliance with a company's design guidelines.

Continuing this initial study, we have planned the below agendas for our follow-up work:
\begin{enumerate}
    \item Making improvements to our pipeline and prompt structure by extending \textit{UI grammar} with each rule's occurrence probability; 
    \item Integrating reasoning steps for the target user, information to display, and supported actions of a UI through Chain-of-Thought prompting \cite{wei2022chain}, a workflow resembling the one of human UI designers;
    \item Conducting multi-faceted layout generation assessments involving human evaluators and more quantitative metrics (e.g. Fréchet inception distance) at a larger scale, to ensure the robustness and applicability of our models;
    \item Experiementing the feasibility of generating high-fidelity UI prototypes with LLMs, as discussed above, and potentially build interactive design-support tools to speed up UI prototyping.
\end{enumerate}


\section{Conclusion}
\label{conclusion}
In this work, we explored Large Language Models' ability to generate mobile user interface layouts through 1-shot, in-context learning. We proposed \textit{UI grammar}, a novel approach to represent the hierarchical relationship between UI elements, and incorporated it into our prompts to steer UI layout generation. Our preliminary results demonstrated LLMs' capabilities to generate high-quality UI layouts with competitive performance, as well as the usefulness of UI grammar in improving certain aspects of generation qualities. We conclude by discussing the implications of using LLMs and UI grammar for future research.

\bibliography{main}
\bibliographystyle{icml2021}

\end{document}